\documentclass[preprint]{aastex631}

\begin{document}

\title{Parallel Diffusion Coefficient of Energetic Charged Particles in the Inner Heliosphere from the Turbulent Magnetic Fields Measured by Parker Solar Probe}

\author[0000-0003-2865-1772]{Xiaohang Chen}
\author[0000-0002-0850-4233]{Joe Giacalone}
\affiliation{Lunar and Planetary Laboratory, University of Arizona, Tucson, AZ United States}
\author[0000-0003-4315-3755]{Fan Guo}
\affiliation{Los Alamos National Laboratory, Theoretical Division, Los Alamos, NM United States}
\author[0000-0001-6038-1923]{Kristopher G. Klein}
\affiliation{Lunar and Planetary Laboratory, University of Arizona, Tucson, AZ United States}

\correspondingauthor{Xiaohang Chen}
\email{xiaohang@lpl.arizona.edu}

\begin{abstract}
Diffusion coefficients of energetic charged particles in turbulent magnetic fields are a fundamental aspect of diffusive transport theory but remain incompletely understood. In this work, we use quasi-linear theory to evaluate the spatial variation of the parallel diffusion coefficient $\kappa_\parallel$ from the measured magnetic turbulence power spectra in the inner heliosphere. We consider the magnetic field and plasma velocity measurements from Parker Solar Probe made during Orbits 5-13. The parallel diffusion coefficient is calculated as a function of radial distance from 0.062 to 0.8 AU, and the particle energy from 100 keV to 1GeV. We find that $\kappa_\parallel$ increases exponentially with both heliocentric distance and energy of particles. The fluctuations in $\kappa_\parallel$ are related to the episodes of large-scale magnetic structures in the solar wind. By fitting the results, we also provide an empirical formula of $\kappa_{\parallel}=(5.16\pm1.22) \times 10^{18} \: r^{1.17\pm0.08} \: E^{0.71\pm 0.02} \; (cm^2/s)$ in the inner heliosphere which can be used as a reference in studying the transport and acceleration of solar energetic particles as well as the modulation of cosmic rays. 
\end{abstract}

\section{Introduction} \label{sec:intro}
The propagation of energetic particles in turbulent magnetic fields is a fundamental problem in astrophysics and space physics. The collective motions of the charged particles can often be understood as a diffusive process and typically modeled by solving the Parker equation or focused transport equation \citep{Parker1965,Roelof1969}. The diffusion tensor in the transport equation is to describe the stochastic motions of the charged particles. It is closely related to a wide range of problems such as the propagation of solar energetic particles (SEPs), the diffusive shock acceleration in various environment, and the modulation of cosmic rays in the heliosphere \citep[e.g.,][]{Jokipii1987,Bieber1994,Zank2000,Giacalone2005,Giacalone2015,Guo2010,kong2017,Chen2022}. Understanding the diffusion coefficient is of importance to both understanding fundamental plasma physics and improving space weather prediction. However, the spatial distribution of the diffusion tensor remains poorly understood and requires knowledge about the interplanetary magnetic turbulence. Earlier observations of the solar wind turbulence are usually made at 1 AU or beyond \citep{Bruno2013,Verscharen2019}. 

The diffusion tensor of energetic charged particles can be separated into a symmetric part consisting of components parallel $\kappa_\parallel$ and perpendicular $\kappa_\perp$ to the mean magnetic field, and an asymmetric part $\kappa_A$ describing the large-scale magnetic gradient and curvature drifts \citep{Jokipii1970,Jokipii1977,Isenberg1979}:
\begin{equation}
    \kappa_{ij}=\kappa_\perp\delta_{ij}+(\kappa_\parallel-\kappa_\perp)\frac{B_iB_j}{B^2}+\epsilon_{ijk}\kappa_A\frac{B_k}{B}
    \label{eq:tensor}
\end{equation}
where $B_i$ is the background, or mean magnetic field vector in index notation, $B=|B_i|$ is the mean magnetic field magnitude, $\delta_{ij}$ is the Kronecker delta, and $\epsilon_{ijk}$ is the Levi-Civita symbol. Considerable progress has been achieved over the past several decades towards understanding the diffusion tensor. Quasi-linear theory (QLT) predicts the spatial diffusion coefficients by relating the pitch-angle scattering with the power spectrum of the magnetic field fluctuations \citep{Jokipii1966}. The standard QLT provides estimates of both the parallel diffusion coefficient, $\kappa_{\parallel}$, and the perpendicular diffusion coefficient, $\kappa_\perp$.  For the latter, QLT predicts that $\kappa_\perp$ is proportional to the power of the magnetic turbulence spectrum at zero wavenumber, and is related to field-line random walk \cite[c.f.][]{Jokipii1969}. It has since been realized that this significantly over-estimates $\kappa_\perp$ \citep{Palmer1982,Dwyer1997}. As such, new theories have been developed to account for cross-field diffusion, such as the nonlinear guiding center theory \citep[NLGC;][]{Matthaeus2003,Shalchi2010,Ruffolo2012}. Perpendicular diffusion is still a perplexing topic since the assumptions of the underlying turbulence properties and wave-particle interaction are still under debate. Numerical simulations \citep[e.g.,][]{Giacalone1999} have shown that the ratio between the perpendicular and parallel diffusion coefficients is nearly independent of energy with a value of about $\kappa_{\perp}/\kappa_{\parallel} \sim 2-4\%$ for the turbulence amplitude similar to those observed in the interplanetary magnetic field at 1AU. Additionally, these numerical simulations give values of $\kappa_\parallel$ which are reasonably consistent with the predictions of QLT. It is worth emphasizing that the energy independent $\kappa_{\perp}/\kappa_{\parallel}$ applies specifically to the particles in resonance with the Kolmogorov spectrum in the magnetostatic turbulence from about $\sim MeV$ to $GeV$. For low- or extremely high-energy particles, the ratio of $\kappa_{\perp}/\kappa_{\parallel}$ has shown various energy dependence, likely due to the the nolinear effects in different composite turbulence conditions \citep[e.g.,][]{Shalchi2004,Dundovic2020}.

Considerable work has been invested to understand the modulation of cosmic rays and SEPs in the solar wind with various turbulence transport models \citep[][]{Schlickeiser1998,Zank1998,Lerche2001,Zhang2009,Qin2015,Zank2017,Chhiber2017,Zhao2017}. These studies either use the test particle simulation or diffusive transport theories to estimate the spatial and energy dependence of $\kappa_{\parallel,\perp}$ in the heliosphere under different turbulence conditions, e.g., slab/2D, anisotropic, incompressible or compressive turbulence (see \citet{Engelbrecht2022} for a detailed review). However, the distributions of $\kappa_{\parallel,\perp}$ in these models are still dependent on the assumptions of turbulence properties. On the other hand, {\it In situ} observations provide us good opportunities to study the spatial diffusion coefficients from direct measurements. For example, $\kappa_\parallel$ can be obtained by fitting the observed time-intensity profiles of SEP events with the solution of the transport equation \citep{Palmer1982,Giacalone2015}. But these results are estimated event by event and may vary significantly. Additionally, one can calculate $\kappa_\parallel$ based on QLT and the measured power spectrum density (PSD) of magnetic turbulence. This approach enables us to estimate $\kappa_\parallel$ under various magnetic turbulence conditions along the spacecraft orbits and has been applied in the previous studies to Wind, Helios and Ulysses observations \citep{Droege1993,Droge2003,Erdos2005}.

Determining the diffusion coefficient from the Sun to the Earth space environment is important to understand the acceleration and transport of energetic particles in the inner heliosphere. NASA's Parker Solar Probe (PSP) mission \citep{Fox2016} opened up the possibility to study the turbulence properties \citep[e.g.,][]{Chen2020,Telloni2021,Halekas2023} as well as the diffusion coefficients in the near Sun space and their radial dependence. In this work, we study the spatial distribution of the parallel diffusion coefficients $\kappa_\parallel$ from 0.062AU (13.3 $R_{Sun}$) to about 0.8AU based on QLT \citep{Jokipii1966,Giacalone1999} using the magnetic field and solar wind velocity measurements from 9 orbits of the PSP mission (Orbits 5-13). We find that $\kappa_\parallel$ increases exponentially with heliocentric distance and particle energy. The magnitude of $\kappa_\parallel$ varies remarkably at different radial distances and orbits. Generally, the quieter episodes of solar wind give rise to a greater $\kappa_\parallel$ while the large-scale magnetic structures, such as coronal mass ejections (CMEs), corotating interaction regions (CIRs) and magnetic switchbacks, will enhance the local turbulence and thus lead to a smaller $\kappa_\parallel$. An empirical formula of $\kappa_\parallel$ in the inner heliosphere is provided considering the distribution of $\kappa_\parallel$ over 9 PSP orbits.

\section{Method} \label{sec:method}
In a weakly turbulent plasma, energetic particles are scattered by low-frequency magnetic fluctuations that are in resonance with the particle gyration. This relationship is described by the standard QLT by assuming that the distribution function and magnetic field can be expressed as the sum of an averaged component and a perturbation. Under this quasi-linear approximation, the spatial diffusion coefficient parallel to the mean magnetic field $\kappa_\parallel$ can be expressed in terms of the pitch-angle diffusion coefficient $D_{\mu \mu}$ as:
\begin{equation}
    \kappa_{\parallel}=\frac{v^2}{4}\int^1_{\mu_{min}} \frac{(1-\mu^2)^2 d\mu}{D_{\mu \mu}}
    \label{eq:Kappa}
\end{equation}
where $v$ is the particle velocity; $\mu$ is the cosine of the pitch angle and $\mu_{min}$ is the minimum value of $\mu$ \citep{Jokipii1966,Roelof1968,Earl1973,Luhmann1976}. $D_{\mu \mu}$ is related to the power spectral density of magnetic fluctuations in the directions transverse to the mean magnetic field:
\begin{equation}
    D_{\mu \mu} = \frac{\pi}{4}\Omega_0(1-\mu^2)\frac{f_{res}P(f_{res})}{B_0^2}
    \label{eq:Dmumu}
\end{equation}
where $B_0$ is the magnitude of magnetic field; $\Omega_0=qB_0/mc$ is the gyrofrequency and $P(f_{res})$ is the value of spectral density at the resonant frequency $f_{res}$. Note that Equation \ref{eq:Dmumu} is only applicable to magnetostatic fluctuations and does not include the dynamical effects which have shown to be important for low-energy particles, especially electrons \citep{Bieber1994,Teufel2003}. In this work, we consider the parallel diffusion coefficient of particles with energy above $\sim 100keV$. The Taylor’s frozen-in hypothesis \citep[hereafter TH;][]{Taylor1938} has been used here to convert the wavenumber $k_{res}$ dependence to the frequency $f_{res}$ dependence through $f_{res} =  k_{res} V_{SW}/2\pi$ where $V_{SW}$ is the magnitude of solar wind velocity and the resonant wavenumber is given by $k_{res}=|\Omega_0/v\mu|$. It's worth noting that TH is only valid when the solar wind velocity is much larger than the Alfv\'en speed but this condition no longer holds in the region near the Sun, especially near the Alfve\'n surface \citep{Kasper2021}. However, near the perihelion, PSP is moving nearly perpendicular to the average local magnetic field and the TH remains a reasonable approximation if the angle between the magnetic field and the spacecraft velocity in the plasma frame is greater than $30^\circ$ \citep{Klein2015,Perez2021}. Generally, TH provides a good approximation beyond 0.1 AU \citep{Chhiber2017} and this covers most of the regions focused on in this work. 

\begin{figure}[ht]
\centering
\includegraphics[width=1\textwidth]{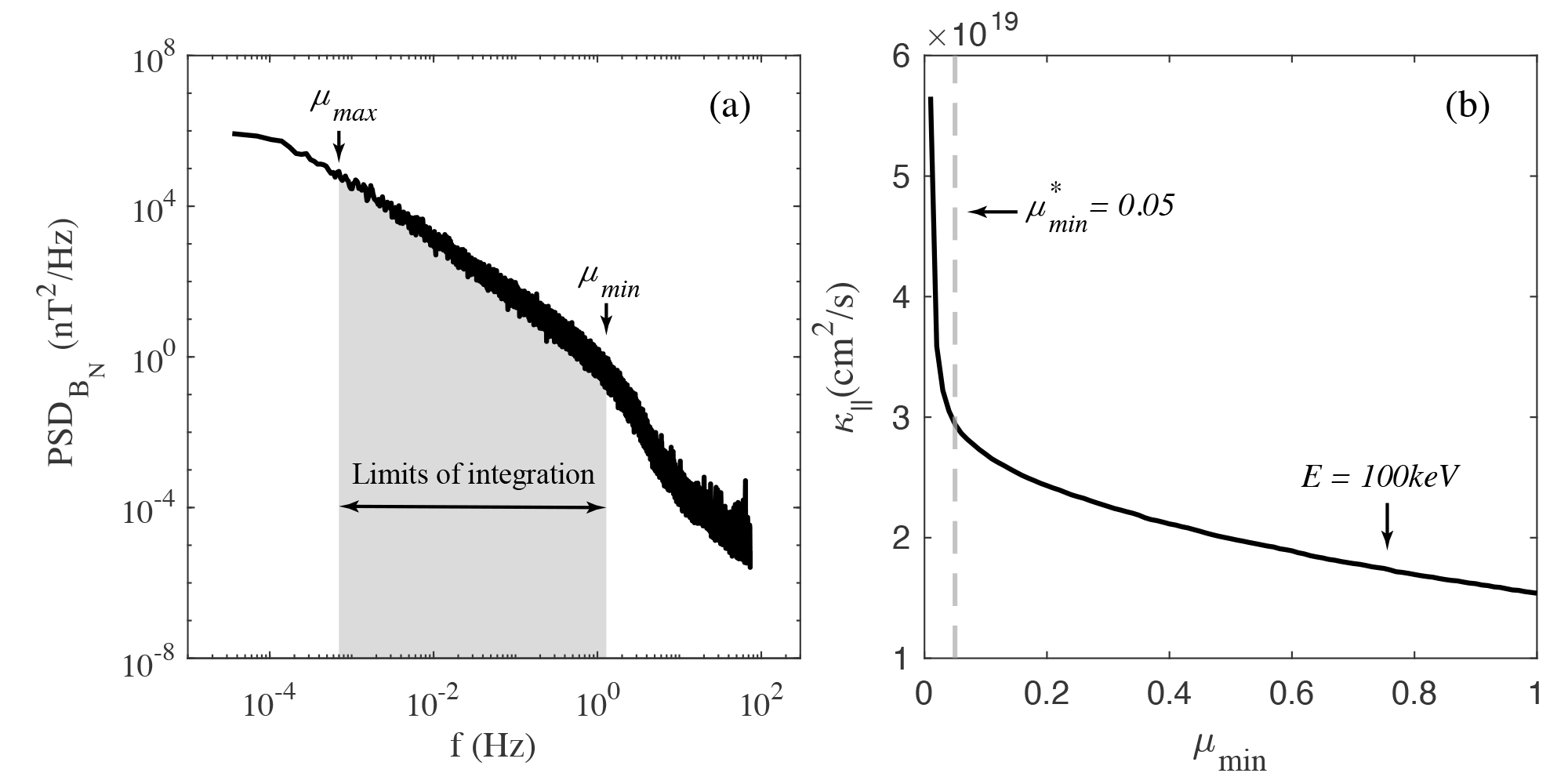}
\caption{(a) Power spectrum density (PSD) of $B_N$ at 0.2AU averaged over Orbit 5-13. The shaded region indicates the range of PSD used to compute $\kappa_\parallel$ in Equation \ref{eq:Kappa} and \ref{eq:Dmumu} with $\mu$ varying from $\mu_{min}$ to $1$. (b) $\kappa_\parallel$ as a function of $\mu_{min}$ in Equation \ref{eq:Kappa}. The solid line is the magnitude of  $\kappa_\parallel$ for $100keV$ protons obtained from the PSD on the left. Here, we consider $\mu_{min}^*=0.05$ (dashed line) to avoid the $90^\circ$ pitch angle scattering problem when $\mu \rightarrow 0$.} 
\label{fig:figure1}
\end{figure}

Thus, $\kappa_\parallel$ can be calculated from the in-situ measurements of $B$ and $V_{SW}$ through Equation \ref{eq:Kappa}-\ref{eq:Dmumu}. We consider the PSP observations during Orbit 5-13 with a radial distance range from 13.3 $R_{Sun}$ to about 0.8 AU. The instrument suites of Solar Wind Electrons, Alphas, and Protons \citep[SWEAP;][]{Kasper2016} and FIELDS \citep{Bale2016} on board PSP provide the data used in this work. Particularly, we use the magnetic field data (in RTN coordinates) from the flux gate magnetometer (MAG) and select the best Solar Probe Analyzer for Ions \citep[SPAN-I;][]{Livi2022}, and Solar Probe Cup \citep[SPC;][]{Case2020} data to calculate the averaged plasma velocity. The normal component $B_N$ in the RTN coordinate system is used to compute the PSD of magnetic turbulence because $B_N$ is always perpendicular to the mean magnetic field (Parker spiral). 
We consider the magnetic field and plasma velocity observations of six-hour intervals, which are sufficiently long to cover the lowest resonant frequency of 1 GeV proton at the aphelion of each orbit. Figure \ref{fig:figure1}(a) presents the averaged power spectrum density of $B_N$ component at 0.2 AU. The shaded region shows the range of PSD used to compute $\kappa_\parallel$ in Equation \ref{eq:Kappa} and \ref{eq:Dmumu} with the pitch cosine $\mu$ varying from $\mu_{min}$ to $1$. We note that the power spectrum shown in Figure \ref{fig:figure1}(a) falls off sharply in high-frequency range. This is due to the dissipation of turbulence at small scales. The dissipation of magnetic turbulence actually causes a well-known problem in the application of QLT in that particles are unable to scatter through $90^\circ$ pitch angle. Figure \ref{fig:figure1}(b) shows how the magnitude of $\kappa_\parallel$ changes with the choice of $\mu_{min}$ in Equation \ref{eq:Kappa}. We can find that $\kappa_\parallel$ blows up at $\mu_{min} \rightarrow 0$ due to this $90^\circ$ pitch angle scattering problem. However, the power spectrum in resonance with particles near $\mu = 0$ corresponds to very high-frequency waves ($f_{res} \rightarrow \infty$) which in fact have little contribution on scattering the high-energy particles (e.g., $100keV$ to $1GeV$ in this paper). Numerical simulations of particles interacting with magnetic turbulence for which the power spectrum falls to zero at high wavenumbers, have revealed that QLT provides a good estimate of the numerically computed spatial diffusion coefficient parallel to the magnetic field \citep{Giacalone1999}. This may be due to the presence of small-magnitudes fluctuations in the magnetic field which permit the mirroring of particles across $\mu=0$. Therefore, we can avoid the non-integrable singularity at $\mu=0$ in Equation \ref{eq:Kappa}  and provide a reasonable estimation of $\kappa_\parallel$ by limiting the minimum value of pitch cosine $\mu_{min}$. In this work, we consider $\mu_{min}^*=0.05$ shown as the dashed line in Figure \ref{fig:figure1}(b).

\section{Results} \label{sec:results}

\begin{figure}[ht]
\centering
\includegraphics[scale=1]{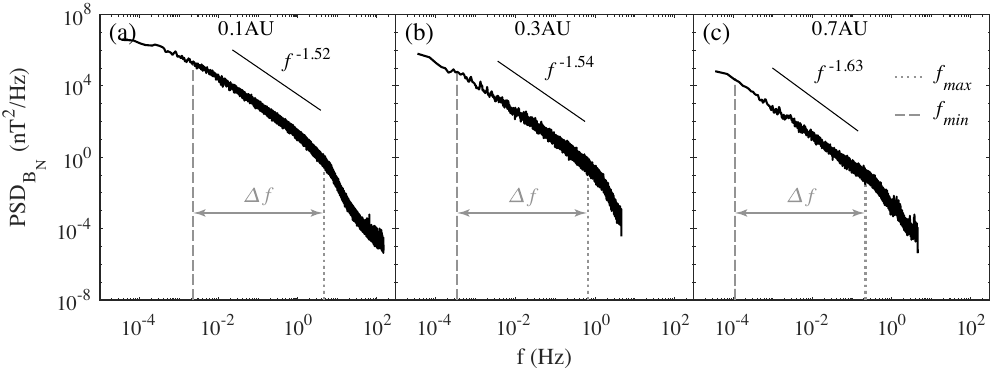}
\caption{PSD of $B_N$ at (a) 0.1 AU, (b) 0.3 AU and (c) 0.7 AU averaged over Orbit 5-13. $\Delta f$ bounded by the minimum and maximum frequencies ($f_{min}$ $\&$ $f_{max}$) indicates the ranges of PSD applied to the calculation of $\kappa_\parallel$ at different heliocentric distances.} 
\label{fig:figure2}
\end{figure}

The parallel diffusion coefficient $\kappa_\parallel$ evolves with the magnetic turbulence in the radial distance from the Sun. Figure \ref{fig:figure2} shows the PSD of six-hour $B_N$ measurements from PSP at (a) 0.1, (b) 0.3 and (c) 0.7 $AU$ averaged over Orbit 5-13. The minimum and maximum frequencies ($f_{min}$ $\&$ $f_{max}$) correspond to the resonant frequencies $f_{res}$ of $1GeV$ protons at $\mu=1$ and $100keV$ protons at $\mu=\mu_{min}$ respectively. We can find that for the energies considered in this work ($100 keV- 1GeV$), the range of the resonant frequency $\Delta f$ is well within the inertial range of the magnetic turbulence. The spectral indices inside $\Delta f$ vary with the heliocentric distance from $-3/2$ to $-5/3$ which are consistent with recent PSP observations \citep[e.g.,][]{Chen2020}. 

\begin{figure}[ht]
\centering
\includegraphics[scale=1]{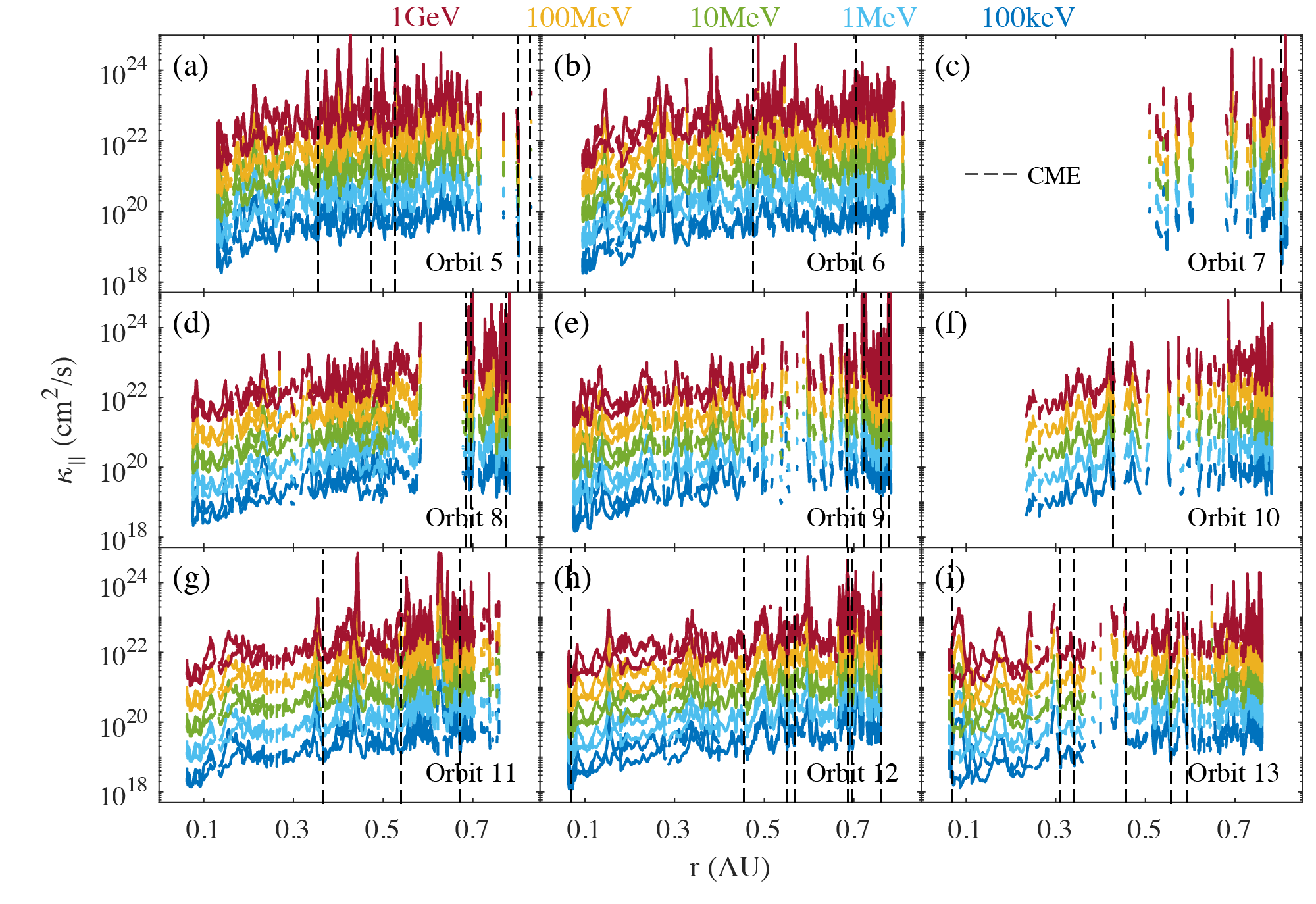}
\caption{(a-i) The parallel diffusion coefficient $\kappa_\parallel$ as a function of radial distance of Orbit 5-13. The dashed vertical lines indicate the onsets of CMEs observed by PSP.}
\label{fig:figure3}
\end{figure}

Figure \ref{fig:figure3}(a)-(i) presents $\kappa_\parallel$ as a function of radial distance from Orbit 5 to 13 with the colors indicating the energy of particles. The blank areas are due to the lack or poor quality of magnetic field or plasma velocity data. We consider the data quality flags for both FIELDS and SWEAP measurements. If any of these flags turn on we just skip that interval. $\kappa_\parallel$ generally increases with heliocentric distance and strong fluctuations can be observed in each orbit. The growth of $\kappa_\parallel$ with $r$ is primarily due to the weakening of the magnetic turbulence during the expansion of the solar wind. The fluctuations in $\kappa_\parallel$ are related to the episodes of magnetic structures. For example, CMEs, CIRs and magnetic switchbacks can enhance the magnetic turbulence and thus lead to a smaller $\kappa_\parallel$. The dashed vertical lines in Figure \ref{fig:figure3} present the CME onsets detected by PSP using the HELIO4CAST ICME catalog \citep{Moestl2023}. A drop of $\kappa_\parallel$ can always be found near the encounter of CMEs. On the other hand, the less turbulent solar wind will give rise to the spikes in Figure \ref{fig:figure3} where particles have a larger mean free path.

\begin{figure}[ht]
\centering
\includegraphics[scale=1]{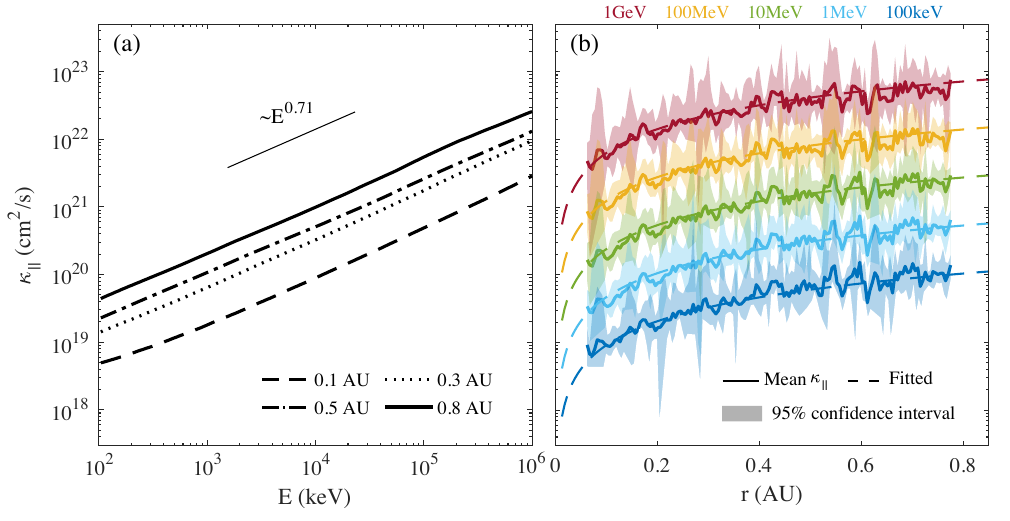}
\caption{(a)The averaged parallel diffusion coefficient $\kappa_\parallel$ as a function of energy. (b) $\kappa_\parallel$ as a function of radial distance. The solid lines are the $\kappa_{\parallel}$ averaged over Orbit 5-13 and the shaded regions indicates the $95\%$ confidence intervals. The dashed lines represents the best fitting results of Equation \ref{eq:kappa_fit}.}
\label{fig:figure4}
\end{figure}

Figure \ref{fig:figure4}(a) examines the averaged parallel diffusion coefficient at several different locations as a function of energy. The calculated kappa is well approximated by a power-law distribution as a function of energy of $\sim 0.71$, close to the expected results $2/3$ from QLT in the turbulence of Kolmogorov spectrum. As mentioned above, this is because the energy range we are interested in has the resonant frequency falling within the inertial range of the turbulence (Figure \ref{fig:figure2}). However, if particle's energy is extremely high, such as cosmic rays, the resonant frequency will enter the low-frequency regions where $\kappa_\parallel$ increases more rapidly with energy and the energy dependence will no longer follow a single power-law distribution and \citep{Dundovic2020,Li2022}. Then, we calculate the mean parallel diffusion coefficient as a function of the radial distance (Figure \ref{fig:figure4}(b)). The solid lines are the mean $\kappa_{\parallel}$ averaged over Orbit 5-13 and the shaded regions indicates the $95\%$ confidence intervals. From the least-squares fitting procedure, we find that the $\kappa_\parallel$ increases as a function of $r$ and energy $E$:
\begin{equation}
    \kappa_{\parallel}=(5.16\pm1.22) \times 10^{18} \: r^{1.17\pm0.08} \: E^{0.71\pm 0.02} \; (cm^2/s)
    \label{eq:kappa_fit}
\end{equation}
where $r$ and $E$ are in the units of $AU$ and $keV$ respectively. The dashed lines in Figure \ref{fig:figure4}(b) show the results of Equation \ref{eq:kappa_fit} in different energies. We observe a gradual increase in the mean value of $\kappa_\parallel$ with radial distance, following a power-law distribution of $r^{1.17\pm0.08}$. Notably, Ulysses also reported a comparable spatial dependence, with $\kappa_\parallel$ scaling as $r^{1.13}$ \citep{Erdos2005}. However, this dependence is still under debate in the inner heliosphere where the interplanetary magnetic field almost radial and scales with $\sim r^{-2}$. Many assumptions of $\kappa \sim r^\gamma$ have been made in SEP transport models with $\gamma$ varying from $0$ to $2$ \citep[e.g.,][]{Zhang2009,Qin2004,Droge2010,Giacalone2020,Wijsen2023}. In our work, this trend is governed by both the magnetic field strength and turbulence magnitudes. For example, the power of magnetic turbulence in resonance with energetic particles (the PSDs within $\Delta f$ between dashed and dotted lines in Figure \ref{fig:figure2}) only changes about 10 times from 0.1 to 0.7 AU while the magnitude of PSD itself drops by more than $10^{3}$ times. Note that this only happens when $\Delta f$ is well falling in the inertial range. Therefore, this empirical formula will provide a useful reference of $\kappa_\parallel$ in the inner heliosphere.

The standard QLT typically predicts $\kappa_\parallel$ to be about order of magnitude smaller than observations, also known as the Palmer consensus \citep{Palmer1982}. To improve the prediction, the anisotropy of the magnetic turbulence is taken into account  \citep[e.g.,][]{Bieber1994,Matthaeus2003,Shalchi2004,Qin2007}.
However, to obtain the turbulence geometry from single-spacecraft observations, a long period of measurements is required to converts the temporal variations into the spatial variations by assuming the solar wind fluctuations are stationary. For example, \citet{Bandyopadhyay2021} considered the five encounters of PSP observations to estimate the 2D correlation function near the Sun. Hence, this approach may not be the optimal choice for resolving the spatial resolution of $\kappa_\parallel$. On the other hand, $\kappa_\parallel$ from test particle simulations have shown a reasonably agreement with the predictions of QLT as noted above. Furthermore, our recent study \citep{Giacalone2023} highlights that QLT offers a relatively accurate estimation for $\kappa_\parallel$ near the CME-driven shocks. These suggest that the current work can also be helpful to the studies of SEP transport in the solar wind as well as acceleration at the interplanetary shocks.  Also, it can be used in global MHD models to promote the reliability of space weather forecasting.

\section{Conclusions} \label{sec:conclusions}
In this paper, we determined the parallel diffusion coefficient $\kappa_\parallel$ in the inner heliosphere as a function of radial distance $r$ from the Sun and the energy $E$ of energetic particles. We applied QLT to calculate $\kappa_\parallel$ based on {\it in-situ} measurements of the turbulent magnetic field and solar wind velocity from Parker Solar Probe for Orbits 5-13. We find that:
\begin{itemize}
  \item $\kappa_\parallel$ increases with $r$ following a power-law distribution: $\kappa_\parallel \propto r^\gamma$ with the best-fitted index $\gamma=1.17$ which is smaller than the assumptions uesd in the previous studies $\gamma=2$.
  \item $\kappa_\parallel$ also increases exponentially with the energy of charged particles: $\kappa_\parallel \propto E^{0.71}$, which is consistent with the prediction from QLT in the Kolmogorov turbulence.
  \item The fluctuations in $\kappa_\parallel$ are associated with the local magnetic structures, such as CMEs, CIRs and magnetic switchbacks.
\end{itemize}
  Finally, We provide an empirical formula of $\kappa_\parallel$ which can be used to infer its value through out the inner heliosphere. Note that Equation \ref{eq:kappa_fit} is only applicable within about $0.1-0.8 AU$ for the energetic particles of the energy between $100keV-1GeV$ since the resonant frequencies $f_{res}$ are generally falling in the inertial range. However, if it is too close to the Sun or the particle has a larger energy $f_{res}$ will enter either the kinetic or low-frequency regions and $\kappa_\parallel$ won't follow a single power law of $r$ and $E$.

\begin{acknowledgments}

We are grateful for helpful discussions with Jung-Tsung Li, Stuart D. Bale, Mihailo Martinovi\'c, Federico Fraschetti and Edmond C. Roelof. Parker Solar Probe was designed, built, and is now operated by the Johns Hopkins Applied Physics Laboratory as part of NASA’s Living with a Star (LWS) program (contract NNN06AA01C). Support from the LWS management and technical team has played a critical role in the success of the Parker Solar Probe mission. We acknowledge PSP/FIELDS team (PI: Stuart D. Bale, UC Berkeley) and PSP/SWEAP team (PI: Justin Kasper, BWX Technologies) for use of data. The project was supported in part by NASA under grants 80NSSC20K1815 and 80HQTR21T0005. X.C. and J.G. acknowledge support from the IS$\odot$IS instrument suite on NASAs Parker Solar Probe Mission, contract NNN06AA01C. F.G. also acknowledges support in part by NASA grant 80HQTR21T0005, 80HQTR21T0087, 80HQTR21T0117, 80HQTR21T0104, and NNH230B17A. 

\end{acknowledgments}

\bibliography{reference}
\end{document}